**Optical properties and corrosion resistance of $Ti_2AlC$, $Ti_3AlC_2$, and $Cr_2AlC$ as candidates for Concentrated Solar Power receivers**

Clio Azina[1,*], Sylvain Badie[2], Andrey Litnovsky[3], Laura Silvestroni[4], Elisa Sani[5], Jesus Gonzalez-Julian[2,6]

[1] Materials Chemistry, RWTH Aachen University, 52074, Aachen, Germany
[2] Institute of Energy and Climate Research: Materials Synthesis and Processing (IEK-1), Forschungszentrum Jülich, 52425 Jülich, Germany
[3] Institute of Energy and Climate Research: Plasma Physics (IEK-4), Forschungszentrum Jülich, 52425 Jülich, Germany
[4] CNR-ISSMC Institute of Science, Technology, and Sustainability for Ceramics, 48018 Faenza, Italy
[5] CNR-INO National Institute of Optics, I-50125 Firenze, Italy
[6] Chair of Ceramics, Institute of Mineral Engineering, RWTH Aachen University, 52074 Aachen, Germany

*Corresponding author; E-Mail: azina@mch.rwth-aachen.de

**Abstract**

New generation concentrated solar power (CSP) plants require new solar receiver materials with selective optical properties and excellent corrosion resistance against molten salts. MAX phases are promising materials for CSP applications due to their optical properties and resistance to thermal shocks. Herein, we report a solar absorptance $\geq 0.5$ and a thermal emittance of 0.17-0.31 between 600 and 1500 K for $Cr_2AlC$, $Ti_2AlC$, and $Ti_3AlC_2$. These compositions were also exposed to solar salt corrosion at 600 °C for up to 4 weeks. $Cr_2AlC$ exhibited superior corrosion resistance due to the formation of a protective nanometric layer.

**1. Introduction**

Concentrated solar power (CSP) is attracting major attention as we are progressing towards more sustainable, safer, and cost-effective technologies for electricity production.

The efficiency of CSP systems is rapidly increasing with operation temperature, requiring materials with good thermal stability, oxidation resistance, and preserved spectral selectivity, to name a few. In that regard, two different CSP receiver concepts have been developed to increase the operating temperature, where either molten salt or air is used as heat transfer fluid. In the first case, the corrosion resistance is the main limitation, whereas in the second it is the oxidation resistance and optical response. Recently, ultra-high temperature ceramics (UHTCs), and in particular diborides [1–4], carbides [3,5], and carbide-based nanocomposites [6,7] have been suggested as suitable alternatives to the more conventional graphite [8] absorbers. Other promising materials which fulfill the specification requirements and which have yet to be fully investigated are MAX phases [9–12]. These ceramics provide a unique combination of metallic and ceramic properties, which can be further tailored by designing solid solutions [13–15], and which allow for excellent oxidation and corrosion resistance, among others [16–22].

Recently, Barua *et al.* showed that Ti-based MAX phases would extend the operating time of the receivers for Gen3 CSP towers as their creep resistance at high temperatures surpasses that of nickel-based alloy structural materials [23]. Furthermore, Sarwar *et al.* carried out studies on $Cr_2AlC$ and $Ti_2AlC$ under high-flux simulator and assessed the promising nature of both MAX phases [24], while Fang *et al.* discussed the potential of $Ti_3SiC_2$ for high-temperature solar absorbers based on their optical performance and thermal stability [25]. However, one of the most critical factors for CSP materials is the hot corrosion resistance against molten salt. Van Loo *et al.* investigated the interaction of different SiC grades, alloys, and MAX phases with $KNO_3$-$NaNO_3$ at 600 °C [26]. $Cr_2AlC$ exhibited the best performance of all tested compositions due to the formation of a stable and sub-micrometer-thick $Cr_7C_3$ scale, which strongly minimized the interaction with the salt [26]. Additional studies are needed, but these preliminary works have shown high potential for MAX phases to increase the efficiency and operating lifetime of CSP systems.

Herein, we studied three MAX phases, $Ti_2AlC$, $Ti_3AlC_2$, and $Cr_2AlC$, which were considered promising based on previous outcomes [24,26]. The optical properties of the as-prepared MAX phases were investigated to compare their performance with that of selected diborides and SiC, while their corrosion resistance in solar salt at 600 °C was assessed for up to 672h.

## 2. Materials and methods

MAX phase powders were prepared by the molten salt shielded synthesis process ($Ti_2AlC$)[27], and by solid-state reaction ($Ti_3AlC_2$, $Cr_2AlC$) [28]. The powders were placed in a graphite mold (30 mm Ø) and sintered in vacuum (<3Pa) by Field Assisted Sintering Technology/Spark Plasma Sintering (FAST/SPS) at 1200 °C with 100 K/min heating rate, for 10 min dwell time, and under 50 MPa uniaxial load. The produced discs were then ground to remove the protective graphite foil, followed by fine grinding with P4000 SiC paper.

The average surface roughness of each sample was determined in the center of each disc using a Contour GT-K 3D non-contact profilometer (Bruker, Germany) on areas of 6×6 $mm^2$ and the data were analyzed using commercial software (Vision64 Map). Optical reflectance spectra at room temperature and quasi-normal incidence angle in the 0.25-2.5 µm wavelength region were acquired using a double-beam spectrophotometer (Perkin Elmer Lambda900) equipped with a 150 mm in diameter integration sphere for the measurement of the hemispherical reflectance. The spectra in the 2.5-15.5 µm wavelength region were acquired using a Fourier Transform spectrophotometer (FT-IR

Bio-Rad Excalibur) equipped with a gold-coated integrating sphere and a liquid nitrogen-cooled detector.

Following the approach adopted in corrosion tests of $Al_2O_3$-forming steel [29] and Ti-based MAX phases [18,30–32], the Ti-based samples were pre-oxidized in air at 1100 °C for 12h (10 K/min), to form dense $Al_2O_3$ scales of ~2.2 and ~1.9 µm in thickness, respectively. The corrosion tests were performed, on ca. 6.5×6.5×4.1 $mm^3$ blocks, by mixing $NaNO_3$ (98+%, Alfa Aesar) and $KNO_3$ (99%, Alfa Aesar) in a 60:40 weight ratio, corresponding to the eutectic molten salt mixture (referenced hereafter as "solar salt"). The blocks were corroded for up to 4 weeks, at 600 °C (10 K/min). The samples were removed from the solar salt after 24, 168, 336, and 672h, and further characterized.

X-ray diffraction (XRD) patterns were collected using a D4 Endeavor (Bruker AXS GmbH) system with Cu radiation. The measurements were carried out over the $2\theta$ range of 5-80°, with a step size of 0.02°. The microstructure was evaluated by scanning electron microscopy (SEM, Zeiss Crossbeam 540, Carl Zeiss AG, Germany) before and after corrosion tests. Scanning transmission electron microscopy (STEM) was carried out on thin lamellae prepared in a FEI Helios NanoLab dual-beam focused ion beam (FIB) microscope, using $Ga^+$ ions accelerated at 30 kV. A STEM III detector was used for imaging. Line scans were collected using an EDAX Octane Elect EDX detector, using an accelerating voltage of 12 kV.

### 3. Results and discussion

Table 1 gathers the density, phases identified by XRD and refined by Rietveld analysis, and the average surface roughness, $R_a$ of the polished as-sintered MAX phases. Sintering led to fully dense samples, achieving relative density values above 99%, as supported by the SEM micrographs in Fig. 1, where almost no porosity is observed. The light grey phases detected in the back-scattered SEM micrographs correspond to the MAX phase, while the darker grey phases are Al$_2$O$_3$ particles as evidenced by the arrows in Fig. 1. Al$_2$O$_3$ was detected in all the MAX phase compositions, which was correlated to the inherent surface oxidation of the starting Al powder used as reactant. Trace amounts of Fe-based compounds were found adjacent to Al$_2$O$_3$ in Ti$_3$AlC$_2$ (Fig.1(b)), which were linked to the milling process. Cr$_2$AlC samples exhibited high purity, presenting just 3 wt.% of Al$_2$O$_3$ and 1 wt.% of Cr$_7$C$_3$ (Fig. 1(c)). While the average surface roughness was in the same order of magnitude for the three MAX phases, Cr$_2$AlC appeared to have the lowest roughness, possibly because of the presence of less secondary phases and therefore less pull out caused by polishing.

Table 1: Relative density, phase composition determined by Rietveld refinement, and average roughness (Ra) of the polished as-sintered MAX phase samples.

| Material | Relative density (%) | Crystalline phase contents (wt.%) | $R_a$ (nm) |
|---|---|---|---|
| Ti$_2$AlC | >99 | Ti$_2$AlC (84%), Ti$_3$AlC$_2$ (14%), Al$_2$O$_3$ (2%) | 12 ± 2 |
| Ti$_3$AlC$_2$ | >99 | Ti$_3$AlC$_2$ (93%), Ti$_2$AlC (2%), Al$_2$O$_3$ (5%) | 27 ± 4 |
| Cr$_2$AlC | >99 | Cr$_2$AlC (96%), Al$_2$O$_3$ (3%), Cr$_7$C$_3$ (1%) | 8 ± 1 |

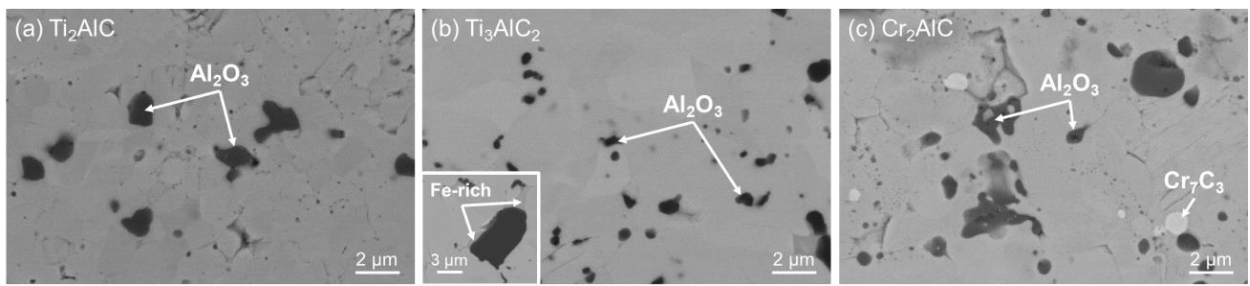

Figure 1: Microstructure of the polished (a) Ti$_2$AlC, (b) Ti$_3$AlC$_2$, and (c) Cr$_2$AlC MAX phases.

On the same mirror-polished surfaces, hemispherical reflectance spectra were measured in the full spectral range 0.3-16.0 µm (Fig. 2(a)) and in the solar spectrum region, 0.3-3.0 µm wavelength (Fig. 2(b)). All three MAX phases showed step-like spectra indicating intrinsic spectral selectivity. The $Ti_2AlC$ and $Ti_3AlC_2$ spectra in Fig.2 nearly superimposed each other in the infrared from 3 µm on, while some minor spectral differences can be seen at shorter wavelengths (Fig. 2(b)). $Cr_2AlC$ had a slightly lower reflectance up to about 10 µm and a curve risefront slightly blueshifted with respect to the Ti-based samples.

From hemispherical reflectance spectra $\rho^\cap(\lambda)$, it is possible to estimate the total hemispherical emittance, $\varepsilon$, at temperature $T$ and the total solar absorptance, $\alpha$, according to the following equations:

$$\varepsilon = \frac{\int_{\lambda_1}^{\lambda_2}(1-\rho^\cap(\lambda))\cdot B(\lambda,T)d\lambda}{\int_{\lambda_1}^{\lambda_2} B(\lambda,T)d\lambda} \qquad (1)$$

where $B(\lambda,T)$ is the blackbody spectral radiance at the temperature $T$ of interest and the integration bounds are $\lambda_1 = 0.3$ µm and $\lambda_2 = 16.0$ µm.

Similarly, the total solar absorptance can be calculated:

$$\alpha = \frac{\int_{\lambda_{min}}^{\lambda_{max}}(1-\rho^\cap(\lambda))\cdot S(\lambda)d\lambda}{\int_{\lambda_{min}}^{\lambda_{max}} S(\lambda)d\lambda} \qquad (2)$$

where $S(\lambda)$ is the A.M. 1.5 sunlight spectral distribution[33] and the integration is carried out between $\lambda_{min}=0.3$ µm and $\lambda_{max}=3.0$ µm.

The two main optical parameters to consider for solar applications are the solar absorptance, $\alpha$, and the $\alpha/\varepsilon$ ratio or spectral selectivity. In an ideal spectrally-selective material, $\alpha$ should be the nearest as possible to unity, while $\alpha/\varepsilon$ should be taken as high as possible. Non-selective (so-called grey) materials, like silicon carbide (SiC), show $\alpha/\varepsilon$ values near to unity. In the case of MAX phases, the $\alpha/\varepsilon$ values have only been reported for $Ti_3SiC_2$ [25], to the best of the authors' knowledge.

Fig. 2(c) shows the temperature-dependent emittance values calculated for the MAX phase samples compared to those of a fully dense SiC reference and transition metal borides [4,34]. As a methodology comment, it should be noticed that the ε and α values given from Eqs. 1-2, being calculated from room temperature spectra, represent an estimation widely used in the literature for a comparative evaluation among materials, while they cannot represent exact values in operative conditions, which would need the knowledge of the spectra at the considered temperature, and result generally underestimated [35]. The values of solar absorptance are indicated in the legend of Fig. 2(c), while the temperature-dependent spectral selectivity values are plotted in Fig. 2(d). The three MAX phases showed similar properties, with α values ≥0.5 and ε in the 0.17-0.31 range at the considered temperatures. The sample with the highest solar absorptance and emittance was $Cr_2AlC$, corresponding to a slightly lower spectral selectivity than the Ti-containing compounds for temperatures below 1100 K. At higher temperatures, the three samples were equivalent.

As for the comparison with previously investigated borides and SiC, the latter being currently the most advanced absorber used in solar plants, MAX phases are significantly more spectrally selective than SiC, and, like borides, have a lower solar absorptance.

However, as we previously reported, this weakness can be mitigated with a selective increase of solar absorptance through proper surface texturing techniques [36,37].

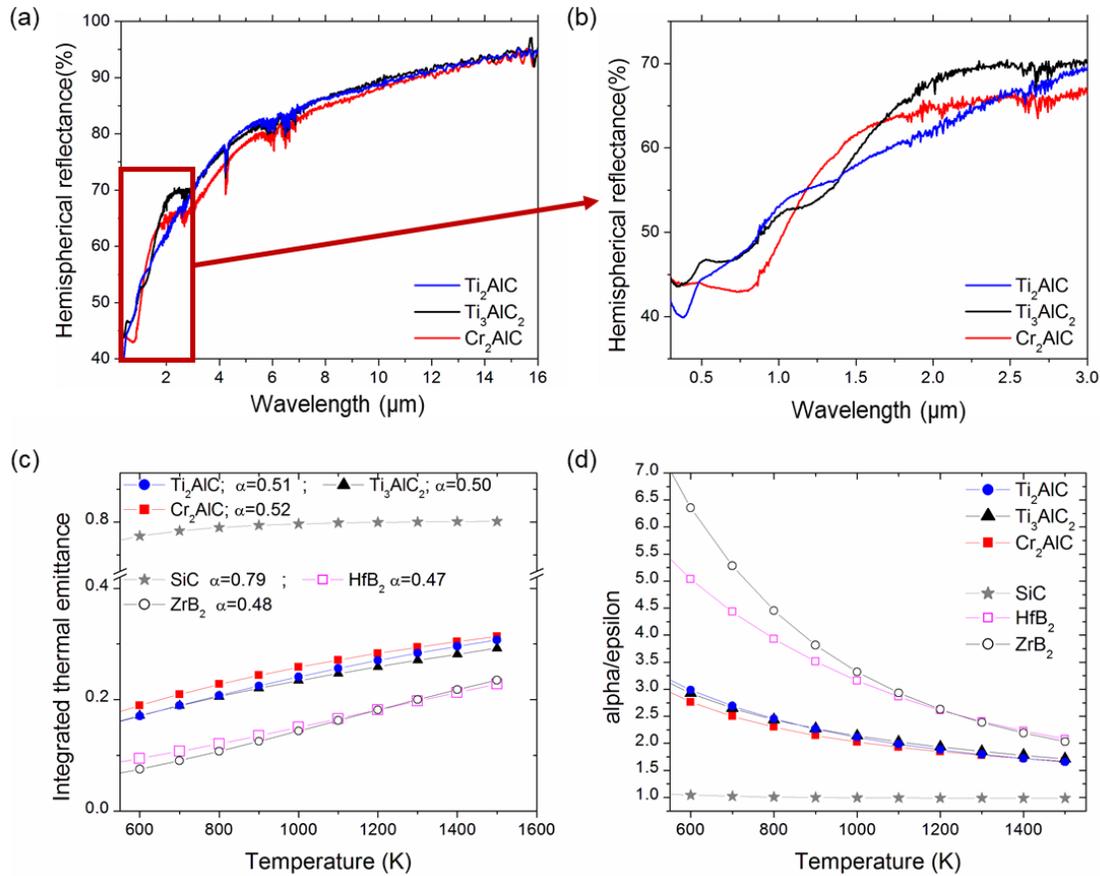

Figure 2: (a) Hemispherical reflectance spectra of MAX samples in the 0.3-16 µm range and (b) detail of the solar spectrum range. (c) Calculated total emittance and (d) calculated spectral selectivity in the 600-1500 K temperature range. Data for SiC, $HfB_2$ and $ZrB_2$ reference materials were taken from [34] and [4]. For a fair comparison, since surface finishing and possible porosity have an impact on the resulting optical properties, the materials were all dense and polished.

The optical properties of the different MAX phases showed high potential for using them in CSP applications, in particular for volumetric air receivers. However, for molten salt CSP systems, corrosion resistance at high temperature is critical. The three MAX phases were corroded by solar salt at 600 ºC for up to 672h, followed by microstructural characterization. The XRD patterns of the as-sintered and corroded $Ti_2AlC$ samples, are provided in Fig. 3(a). Peaks corresponding to $Al_2O_3$, $Na_5AlO_4$, $NaTi_2O_4$, and $NaAl_6O_{9.5}$ were detected from 168h, and increased in intensity with corrosion time. Furthermore, one can notice that the MAX phase peaks disappeared almost completely after 168h, hinting at a considerable thickness of corrosion layer, which was then confirmed by cross-section imaging, as shown in Fig. 3(b)-(d). Indeed, the corrosion layer was inhomogeneous and nearly 10 µm-thick even after 24h. The scale became more homogeneous in thickness after 336h and remained stable at ~10 µm. The scale appeared to be somewhat lamellar and quite porous, suggesting it could perform poorly in terms of corrosion protection. As the corrosion time reached 672h, the thickness of the scale averaged at 20 µm. Chemical mapping performed over the area shown in Fig. 3(e), showed that the scale contained primarily Ti, Na and Ca, whereas Al was detected on the outmost surface.

Overall, these observations led to conclude that the $Al_2O_3$ scale grown prior to the corrosion tests did not yield the expected protective effect. In fact, the alumina layer may have behaved in two different ways. First, the oxide scale could have been consumed by the solar salt which would be supported by observations from Van Loo *et al.* who did not identify $Al_2O_3$ after corrosion of untreated (not pre-oxidized) commercial Maxthal® $Ti_2AlC$ and $Ti_3AlC_2$ samples in the same solar salt for up to 1000h [26]. The second scenario would imply spalling of the alumina scale. Considering that the thickness of the $Al_2O_3$ layer was ~2 µm, the stresses generated during corrosion may have led to the premature fragmentation of the scale, implying that the scale was perhaps too thin [38]. The salt could have, then, easily corroded the underlying MAX phase. This hypothesis seems to be more reasonable as it is supported by the indexing of $Al_2O_3$ by XRD and the Al-rich leftovers seen in Fig. 3(e). Finally, despite the presence of significant amounts of corrosion products, the $Ti_2AlC$ sample exhibited an overall good resistance to the solar salt as it maintained its integrity even after 672h, as illustrated in the inset of Fig. 3(a).

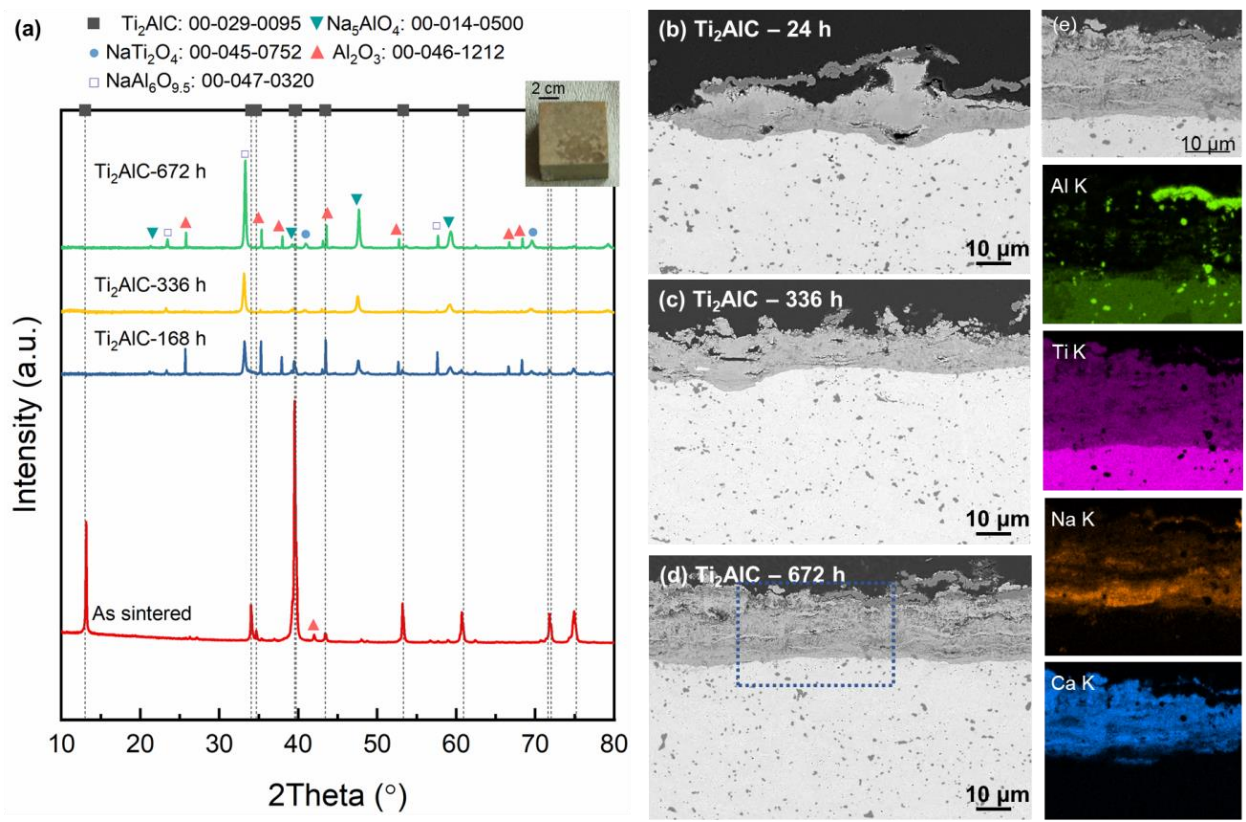

Figure 3: (a) XRD patterns of as-sintered and corroded $Ti_2AlC$ samples. (b)-(d) cross-section micrographs of corroded samples after 24, 336 and 672h. (e) Magnified image and EDS elemental maps of the rectangular zone in (d). The inset in (a) shows the appearance of the corroded 6.5×6.5×4.1 $cm^3$ sample after 672h corrosion.

The behavior of $Ti_3AlC_2$ towards corrosion differed significantly. As it can be seen from the XRD patterns in Fig. 4(a), a multitude of peaks corresponding to various oxides appeared after 168h. In this case, aside from Na-containing aluminides and titanates, $TiO_2$ was also indexed. The poorer corrosion resistance of $Ti_3AlC_2$ was further demonstrated by the degradation of the sample with corrosion time (Fig. 4(b)). The corrosion initiated at the corners and edges, where thermal and mechanical stresses between the pristine sample and corrosion layer are greater. After 672h, the sample completely disintegrated. The cross-section micrograph of the sample corroded for 24h is shown in Fig. 4(c). While the corrosion scale seemed relatively dense, its thickness had already reached ~68 μm, confirming the poor corrosion resistance of $Ti_3AlC_2$ in solar salt. From the EDS maps, one can notice the absence of the continuous $Al_2O_3$ scale formed prior to the corrosion tests, but rather the presence of Al-rich grains which could be leftovers of the intentionally formed $Al_2O_3$ scale. Furthermore, one can notice Ti-rich zones indicating that the composition within the scale was inhomogeneous, also supported by the chemical contrast seen in the scale in Fig. 2(c). In this case, K was also detected in the reaction layer.

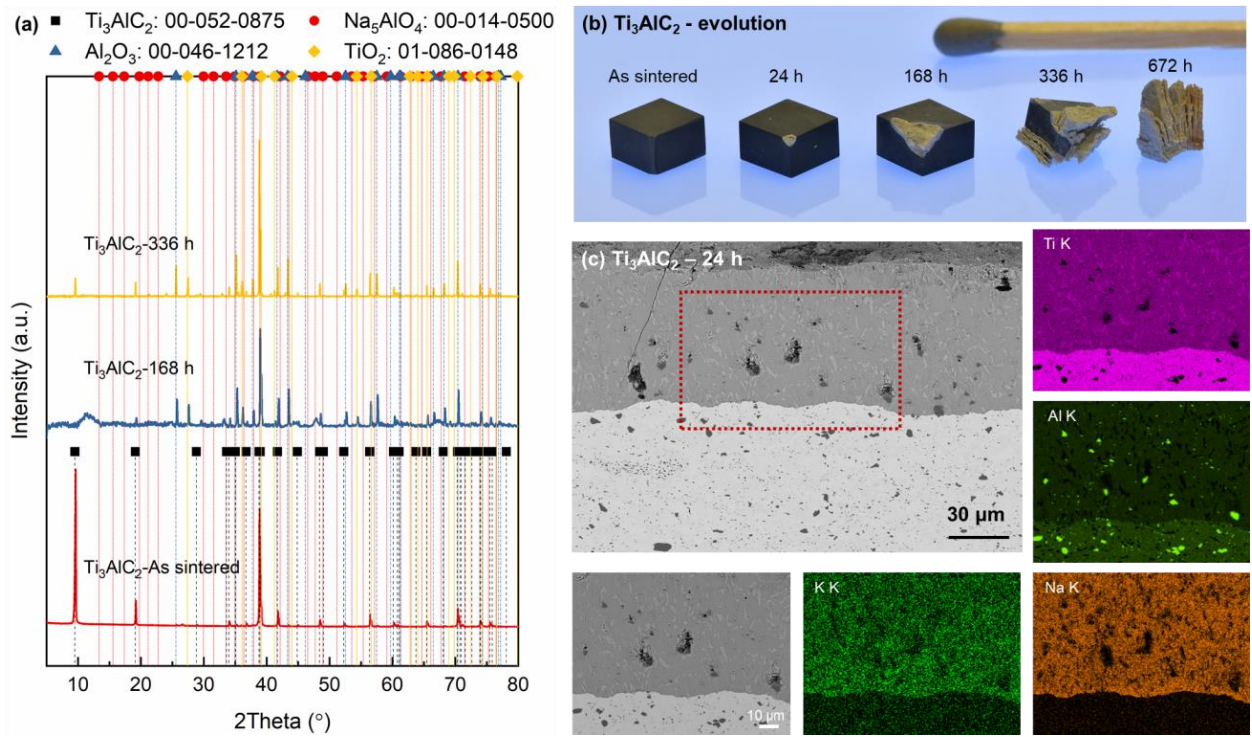

Figure 4: (a) XRD patterns of as-sintered and corroded $Ti_3AlC_2$ samples. (b) Photographs of the samples post-sintering and after 24, 168, 336 and 672h corrosion in solar salt. A match is provided for size reference (c) Cross-section micrograph of the corroded sample after 24h with corresponding EDS elemental maps of the rectangular zone.

$Cr_2AlC$ did not appear to interact with the solar salt even after 672h, as confirmed by both XRD and SEM cross-sections (Fig. 5(a)-(d)), indicating excellent resistance to solar salt, which is in good agreement with observations made by Van Loo *et al.* [26]. The ~245 nm-thick scale formed was dense and appeared homogeneous after 168h (Fig. 5(e,g)).

The EDS line scan in Fig. 5(f) shows that the oxide scale is fairly complex. Cr, Na and C enrichment can be noticed in the outermost surface; however, the thin scale was primarily Al- and O-based. This seems to be in contrast with the observations made for the Ti-based MAX phases. In addition, Mg, Na and Si were also detected in the corrosion layer.

The excellent corrosion resistance of $Cr_2AlC$ could be attributed to the Cr-rich outermost surface. Indeed, the corrosion resistance of a Cr-containing stainless steel was shown to be superior than that of a Cr-free carbon steel as the corrosion scales after 21 days reached 158 and 610 nm, respectively in solar salt at 390 °C [39]. Compositional depth profiles of the steel samples showed that Cr diffused to the surface and its content increased with increasing corrosion time. However, it is difficult at this stage to make any conclusive statement as the chemical environment of Cr in steel differs significantly from that in $Cr_2AlC$. Based on the observations made by Van Loo *et al.*, the possibility of forming a $Cr_7C_3$ outer layer, with no significant oxygen uptake, would explain the enhanced corrosion resistance of $Cr_2AlC$ [26], however, in our case, a significant amount of oxygen was detected in the outermost thin scale (see Fig. 5(e), (f)) suggesting the presence of an oxide phase. Further focused investigation is ongoing to reveal the nature of this thin protective layer, which showed an excellent corrosion resistance towards the solar salt.

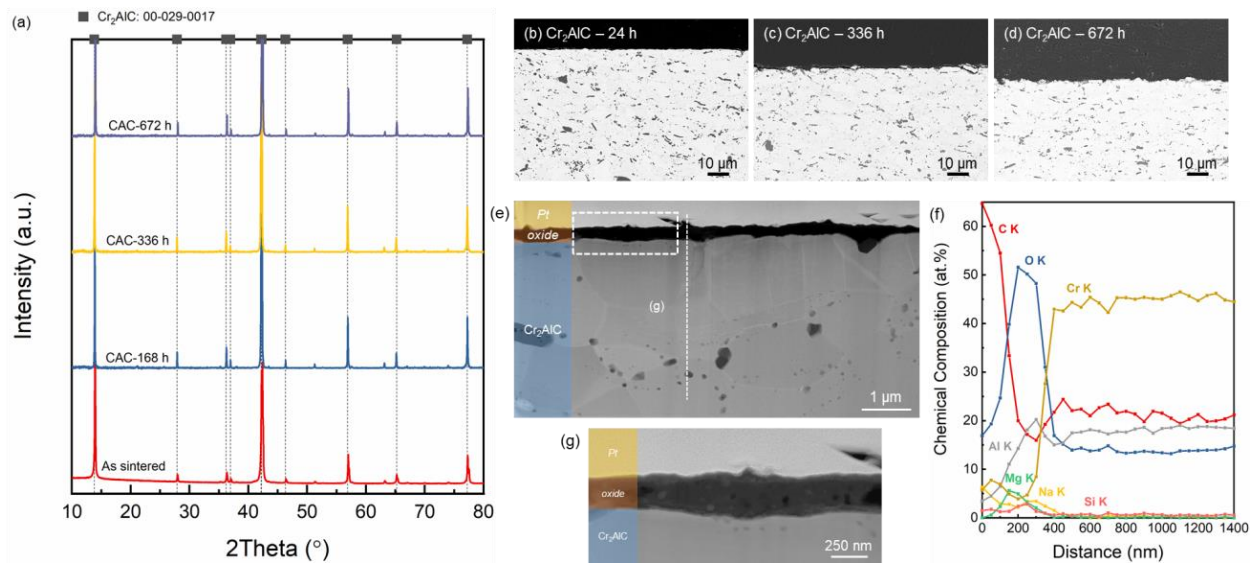

Figure 5: (a) XRD patterns of Cr$_2$AlC samples before and after corrosion in solar salt. SEM cross-section micrographs of Cr$_2$AlC samples corroded for (b) 24, (c) 336 and (d) 672h in solar salt. (e) HAADF STEM image of Cr$_2$AlC corroded for 168h in solar salt, (f) EDS line scan collected along the dashed line in (e), (g) high magnification micrograph of the corrosion layer in HAADF mode corresponding to the dashed frame in (e).

4. **Conclusion**

The optical properties and hot corrosion resistance against solar salt of Ti$_2$AlC, Ti$_3$AlC$_2$ and Cr$_2$AlC MAX phases were investigated to evaluate their potential for use as concentrated solar power receivers. The three materials presented similar optical responses, with solar absorptances $\geq$ 0.5 and thermal emittances in the 0.17-0.31 range between 600 and 1500 K. The solar absorptance values were slightly higher than those of ZrB$_2$ and HfB$_2$, but lower than SiC. Likewise, the thermal emittance and spectral selectivity ($\alpha/\varepsilon$) of MAX phases lied between those of SiC and the borides. Therefore, these three MAX phases present relevant optical properties for further development in CSP systems, in particular for volumetric air receivers where the optical properties are more determinant.

The corrosion resistance of the three MAX phases was also tested in a NaNO$_3$-KNO$_3$ solar salt mixture at 600 °C for up to 4 weeks. Ti$_2$AlC and Cr$_2$AlC performed quite well, while $Ti_3AlC_2$ fully decomposed. In particular, $Cr_2AlC$ exhibited minimal signs of corrosion after 4 weeks. A thin corrosion layer of ~245 nm was formed after 168h, while that of $Ti_2AlC$ was already close to 10 µm-thick after just 24h. While the corrosion layer thicknesses differed, both $Ti_2AlC$ and $Cr_2AlC$ MAX phases exhibited good corrosion resistance in solar salt making them relevant for application as molten-salt receivers and heat exchangers.

**Acknowledgments**

C.A. acknowledges funding from the European Union's H2020-MSCA-IF-2019 research and innovation programme under the Marie Skłodowska-Curie grant agreement No. 892501 (REALMAX). The authors would like to thank Ms Melina Poll for sintering the samples and Prof. Jochen M. Schneider for access to the Focused Ion Beam (FIB).

**Declaration of interest statement**

The authors confirm that there are no known conflicts of interest associated with this publication.

**References**

[1] D. Sciti, L. Silvestroni, J.L. Sans, L. Mercatelli, M. Meucci, E. Sani, Tantalum diboride-based ceramics for bulk solar absorbers, Solar Energy Materials and Solar Cells. 130 (2014) 208–216. https://doi.org/10.1016/j.solmat.2014.07.012.
[2] L. Silvestroni, D. Sciti, L. Zoli, A. Balbo, F. Zanotto, R. Orrù, R. Licheri, C. Musa, L. Mercatelli, E. Sani, An overview of ultra-refractory ceramics for thermodynamic solar energy generation at high temperature, Renewable Energy. 133 (2019) 1257–1267. https://doi.org/10.1016/j.renene.2018.08.036.
[3] E. Sani, L. Mercatelli, J.L. Sans, L. Silvestroni, D. Sciti, Porous and dense hafnium and zirconium ultra-high temperature ceramics for solar receivers, Optical Materials. 36 (2013) 163–168. https://doi.org/10.1016/j.optmat.2013.08.020.
[4] E. Sani, L. Mercatelli, M. Meucci, L. Silvestroni, A. Balbo, D. Sciti, Process and composition dependence of optical properties of zirconium, hafnium and tantalum borides for solar receiver applications, Solar Energy Materials and Solar Cells. 155 (2016) 368–377. https://doi.org/10.1016/j.solmat.2016.06.028.
[5] D. Sciti, L. Silvestroni, D.M. Trucchi, E. Cappelli, S. Orlando, E. Sani, Femtosecond laser treatments to tailor the optical properties of hafnium carbide for solar

applications, Solar Energy Materials and Solar Cells. 132 (2015) 460–466. https://doi.org/10.1016/j.solmat.2014.09.037.

[6] M. Coulibaly, G. Arrachart, A. Mesbah, X. Deschanels, From colloidal precursors to metal carbides nanocomposites MC (M=Ti, Zr, Hf and Si): Synthesis, characterization and optical spectral selectivity studies, Solar Energy Materials and Solar Cells. 143 (2015) 473–479. https://doi.org/10.1016/j.solmat.2015.07.039.

[7] H. Aréna, M. Coulibaly, A. Soum-Glaude, A. Jonchère, G. Arrachart, A. Mesbah, N. Pradeilles, M. Vandenhende, A. Maître, X. Deschanels, Effect of TiC incorporation on the optical properties and oxidation resistance of SiC ceramics, Solar Energy Materials and Solar Cells. 213 (2020) 110536. https://doi.org/10.1016/j.solmat.2020.110536.

[8] C.C. Agrafiotis, I. Mavroidis, A.G. Konstandopoulos, B. Hoffschmidt, P. Stobbe, M. Romero, V. Fernandez-Quero, Evaluation of porous silicon carbide monolithic honeycombs as volumetric receivers/collectors of concentrated solar radiation, Solar Energy Materials and Solar Cells. 91 (2007) 474–488. https://doi.org/10.1016/J.SOLMAT.2006.10.021.

[9] J. Gonzalez-Julian, Processing of MAX phases: From synthesis to applications, Journal of the American Ceramic Society. 104 (2021) 659–690. https://doi.org/10.1111/jace.17544.

[10] P. Eklund, M. Beckers, U. Jansson, H. Högberg, L. Hultman, The $M_{n+1}AX_n$ phases: Materials science and thin-film processing, Thin Solid Films. 518 (2010) 1851–1878. https://doi.org/10.1016/j.tsf.2009.07.184.

[11] M.W. Barsoum, T. El-Raghy, The MAX phases: Unique new carbide and nitride materials, American Scientist. 89 (2001) 334–343. https://doi.org/10.1511/2001.4.334.

[12] M. Radovic, M.W. Barsoum, MAX phases: Bridging the gap between metals and ceramics, American Ceramic Society Bulletin. 92 (2013) 20–27.

[13] C. Azina, T. Bartsch, D.M. Holzapfel, M. Dahlqvist, J. Rosen, L. Löfler, A.S.J. Mendez, M. Hans, D. Primetzhofer, J.M. Schneider, Yttrium incorporation in $Cr_2AlC$: On the metastable phase formation and decomposition of $(Cr,Y)_2AlC$ MAX phase thin films, Journal of the American Ceramic Society. n/a (2022) 1–14. https://doi.org/10.1111/jace.18931.

[14] C. Azina, B. Tunca, A. Petruhins, B. Xin, M. Yildizhan, P.O.Å. Persson, J. Vleugels, K. Lambrinou, J. Rosen, P. Eklund, Deposition of MAX phase-containing thin films from a $(Ti,Zr)_2AlC$ compound target, Applied Surface Science. 551 (2021) 149370. https://doi.org/10.1016/j.apsusc.2021.149370.

[15] B. Tunca, S. Huang, N. Goossens, K. Lambrinou, J. Vleugels, Chemically complex double solid solution MAX phase-based ceramics in the (Ti,Zr,Hf,V,Nb)-(Al,Sn)-C system, Http://Mc.Manuscriptcentral.Com/Tmrl. 10 (2022) 52–61. https://doi.org/10.1080/21663831.2021.2017043.

[16] H. Shi, R. Azmi, L. Han, C. Tang, A. Weisenburger, A. Heinzel, J. Maibach, M. Stüber, K. Wang, G. Müller, Corrosion behavior of Al-containing MAX-phase coatings exposed to oxygen containing molten Pb at 600 °C, Corrosion Science. 201 (2022) 110275. https://doi.org/10.1016/j.corsci.2022.110275.

[17] Z.J. Lin, M.S. Li, J.Y. Wang, Y.C. Zhou, High-temperature oxidation and hot corrosion of $Cr_2AlC$, Acta Materialia. 55 (2007) 6182–6191. https://doi.org/10.1016/j.actamat.2007.07.024.

[18] L. Guo, Z. Yan, X. Wang, Q. He, Ti 2 AlC MAX phase for resistance against CMAS attack to thermal barrier coatings, Ceramics International. 45 (2019) 7627–7634. https://doi.org/10.1016/j.ceramint.2019.01.059.

[19] E. Charalampopoulou, K. Lambrinou, T. Van der Donck, B. Paladino, F. Di Fonzo, C. Azina, P. Eklund, S. Mráz, J.M. Schneider, D. Schryvers, R. Delville, Early stages of dissolution corrosion in 316 L and DIN 1.4970 austenitic stainless steels with and without anticorrosion coatings in static liquid lead-bismuth eutectic (LBE) at 500 °C, Materials Characterization. 178 (2021) 111234. https://doi.org/10.1016/j.matchar.2021.111234.

[20] J. Cao, Z. Yin, H. Li, G. Gao, X. Zhang, Tribological and mechanical properties of Ti2AlC coating at room temperature and 800 °C, Ceramics International. 44 (2018) 1046–1051. https://doi.org/10.1016/j.ceramint.2017.10.045.

[21] D.E. Hajas, M. To Baben, B. Hallstedt, R. Iskandar, J. Mayer, J.M. Schneider, Oxidation of Cr2AlC coatings in the temperature range of 1230 to 1410°C, Surface and Coatings Technology. 206 (2011) 591–598. https://doi.org/10.1016/j.surfcoat.2011.03.086.

[22] J. Gonzalez-Julian, T. Go, D.E. Mack, R. Vaßen, Environmental resistance of Cr2AlC MAX phase under thermal gradient loading using a burner rig, Journal of the American Ceramic Society. 101 (2018) 1841–1846. https://doi.org/10.1111/jace.15425.

[23] B. Barua, M.C. Messner, D. Singh, Assessment of Ti3SiC2 MAX phase as a structural material for high temperature receivers, SOLARPACES 2020: 26th International Conference on Concentrating Solar Power and Chemical Energy Systems. 2445 (2022) 030002. https://doi.org/10.1063/5.0085952.

[24] J. Sarwar, T. Shrouf, A. Srinivasa, H. Gao, M. Radovic, K. Kakosimos, Characterization of thermal performance, flux transmission performance and optical properties of MAX phase materials under concentrated solar irradiation, Solar Energy Materials and Solar Cells. 182 (2018) 76–91. https://doi.org/10.1016/j.solmat.2018.03.018.

[25] Z. Fang, C. Lu, C. Guo, Y. Lu, D. Gao, Y. Ni, J. Kou, Z. Xu, P. Li, Suitability of layered Ti3SiC2 and Zr3[Al(si)]4C6 ceramics as high temperature solar absorbers for solar energy applications, Solar Energy Materials and Solar Cells. 134 (2015) 252–260. https://doi.org/10.1016/J.SOLMAT.2014.12.008.

[26] K. Van Loo, T. Lapauw, N. Ozalp, E. Ström, K. Lambrinou, J. Vleugels, Compatibility of SiC–and MAX phase-based ceramics with a KNO 3 -NaNO 3 molten solar salt, Solar Energy Materials and Solar Cells. 195 (2019) 228–240. https://doi.org/10.1016/j.solmat.2019.03.007.

[27] S. Badie, A. Dash, Y.J. Sohn, R. Vaßen, O. Guillon, J. Gonzalez-Julian, Synthesis, sintering, and effect of surface roughness on oxidation of submicron Ti2AlC ceramics, Journal of the American Ceramic Society. 104 (2021) 1669–1688. https://doi.org/10.1111/jace.17582.

[28] J. Gonzalez-julian, S. Onrubia, M. Bram, C. Broeckmann, R. Vassen, O. Guillon, High-temperature oxidation and compressive strength of Cr 2 AlC MAX phase foams with controlled porosity, Journal of the American Ceramic Society. 101 (2017) 542–552. https://doi.org/10.1111/jace.15224.

[29] F. Aarab, B. Kuhn, A. Bonk, T. Bauer, A New Approach to Low-Cost, Solar Salt-Resistant Structural Materials for Concentrating Solar Power (CSP) and Thermal

Energy Storage (TES), Metals. 11 (2021) 1970. https://doi.org/10.3390/met11121970.

[30] L. Guo, G. Li, J. Wu, X. Wang, Effects of pellet surface roughness and pre-oxidation temperature on CMAS corrosion behavior of Ti2AlC, Journal of Advanced Ceramics. 11 (2022) 945–960. https://doi.org/10.1007/s40145-022-0588-0.

[31] Z. Lin, Y. Zhou, M. Li, J. Wang, Improving the Na2SO4-induced corrosion resistance of Ti3AlC2 by pre-oxidation in air, Corrosion Science. 48 (2006) 3271–3280. https://doi.org/10.1016/j.corsci.2005.11.005.

[32] Z. Lin, Y. Zhou, M. Li, J. Wang, Hot corrosion and protection of Ti2AlC against Na2SO4 salt in air, Journal of the European Ceramic Society. 26 (2006) 3871–3879. https://doi.org/10.1016/j.jeurceramsoc.2005.12.004.

[33] Standard Tables for Reference Solar Spectral Irradiances: Direct Normal and Hemispherical on 37° Tilted Surface, (n.d.). https://www.astm.org/g0173-03r20.html (accessed October 17, 2022).

[34] D. Sciti, L. Silvestroni, L. Mercatelli, J.L. Sans, E. Sani, Suitability of ultra-refractory diboride ceramics as absorbers for solar energy applications, Solar Energy Materials and Solar Cells. 109 (2013) 8–16. https://doi.org/10.1016/j.solmat.2012.10.004.

[35] N. Azzali, M. Meucci, D. Di Rosa, L. Mercatelli, L. Silvestroni, D. Sciti, E. Sani, Spectral emittance of ceramics for high temperature solar receivers, Solar Energy. 222 (2021) 74–83. https://doi.org/10.1016/j.solener.2021.05.019.

[36] E. Sani, D. Sciti, L. Silvestroni, A. Bellucci, S. Orlando, D.M. Trucchi, Tailoring optical properties of surfaces in wide spectral ranges by multi-scale femtosecond-laser texturing: A case-study for TaB2 ceramics, Optical Materials. 109 (2020) 110347. https://doi.org/10.1016/j.optmat.2020.110347.

[37] E. Sani, D. Sciti, S. Failla, C. Melandri, A. Bellucci, S. Orlando, D.M. Trucchi, Multi-Scale Femtosecond-Laser Texturing for Photothermal Efficiency Enhancement on Solar Absorbers Based on TaB2 Ceramics, Nanomaterials. 13 (2023) 1692. https://doi.org/10.3390/nano13101692.

[38] S. Badie, D. Sebold, R. Vaßen, O. Guillon, J. Gonzalez-Julian, Mechanism for breakaway oxidation of the Ti2AlC MAX phase, Acta Materialia. 215 (2021). https://doi.org/10.1016/j.actamat.2021.117025.

[39] F. Pineda, M. Walczak, F. Vilchez, C. Guerra, R. Escobar, M. Sancy, Evolution of corrosion products on {ASTM} {A36} and {AISI} {304L} steels formed in exposure to molten {NaNO3}–{KNO3} eutectic salt: {Electrochemical} study, Corrosion Science. 196 (2022) 110047. https://doi.org/10.1016/j.corsci.2021.110047.